\documentstyle[12pt]{article}

\def\hquad{\;}
\def\pT{ \widehat T }
\def\tS{{\widetilde S}}
\def\tR{{\widetilde R}}
\def\ex{{\rm e}}
\def\nbr{ {2i} }
\def\sign{\sigma}
\def\gauge{ e}
\def\eps{\varepsilon}

\catcode`@=11
\def\citen#1{\if@filesw \immediate\write \@auxout {\string\citation{#1}}\fi%
\@tempcntb\m@ne \let\@h@ld\relax \def\@citea{}%
\@for \@citeb:=#1\do {\@ifundefined {b@\@citeb}%
    {\@h@ld\@citea\@tempcntb\m@ne{\bf ?}%
    \@warning {Citation `\@citeb ' on page \thepage \space undefined}}%
    {\@tempcnta\@tempcntb \advance\@tempcnta\@ne
    \setbox\z@\hbox\bgroup\ifcat0\csname b@\@citeb \endcsname \relax
    \egroup \@tempcntb\number\csname b@\@citeb \endcsname \relax
    \else \egroup \@tempcntb\m@ne \fi \ifnum\@tempcnta=\@tempcntb
    \ifx\@h@ld\relax \edef \@h@ld{\@citea\csname b@\@citeb\endcsname}%
    \else \edef\@h@ld{\hbox{--}\penalty\@highpenalty
    \csname b@\@citeb\endcsname}\fi
    \else \@h@ld\@citea\csname b@\@citeb \endcsname \let\@h@ld\relax \fi}%
\def\@citea{,\penalty\@highpenalty\hskip.13em plus.13em minus.13em}}\@h@ld}
\def\@citex[#1]#2{\@cite{\citen{#2}}{#1}}%
\def\@cite#1#2{\leavevmode\unskip\ifnum\lastpenalty=\z@\penalty\@highpenalty\fi%
   $^{\scriptscriptstyle \multiply\@highpenalty 3 \mbox{\rm\scriptsize#1%
  \if@tempswa,\penalty\@highpenalty\ #2\fi}}$}   %

\def\dcite{\@ifnextchar [{\@tempswatrue\@dcitex}{\@tempswafalse\@dcitex[]}}

\def\@dcitex[#1]#2{\if@filesw\immediate\write\@auxout{\string\citation{#2}}\fi
  \def\@dcitea{}\@dcite{\@for\@dciteb:=#2\do
    {\@dcitea\def\@dcitea{,}\@ifundefined
       {b@\@dciteb}{{\bf ?}\@warning
       {d(line)cite  `\@dciteb' on page \thepage \space undefined}}%
\hbox{\csname b@\@dciteb\endcsname}}}{#1}}

\def\@dcite#1#2{$\mbox{\rm#1\if@tempswa , #2\fi}$}
\catcode`@=12

\def\onward{\addtocounter{section}{1} \setcounter{equation}{0} }

\def\smultab#1#2#3#4#5#6#7{\put (0,4){\line(1,0){#1}}
                    \multiput(0,3)(1,0){#1}{\line(1,0){1}}
                    \multiput(1,3)(1,0){#1}{\line(0,1){1}}
                    \multiput(0,2)(1,0){#2}{\line(1,0){1}}
                    \multiput(1,2)(1,0){#2}{\line(0,1){1}}
                    \multiput(0,1)(1,0){#3}{\line(1,0){1}}
                    \multiput(1,1)(1,0){#3}{\line(0,1){1}}
                    \multiput(0,0)(1,0){#4}{\line(1,0){1}}
                    \multiput(1,0)(1,0){#4}{\line(0,1){1}}
                    \multiput(0,-1)(1,0){#5}{\line(1,0){1}}
                    \multiput(1,-1)(1,0){#5}{\line(0,1){1}}
                    \multiput(0,-2)(1,0){#6}{\line(1,0){1}}
                    \multiput(1,-2)(1,0){#6}{\line(0,1){1}}
                          \put (0,4){\line(0,-1){#7}}}
\def\young#1#2#3#4#5#6#7{\begin{picture}(#1,#7)(0,3)
                \thicklines \smultab#1#2#3#4#5#6#7 \end{picture}}

\def\Sig{\Sigma}
\def\col{k}

\def\hchi{{\raise 0.8mm \hbox{$\chi$}}}
\newcommand{\eq}{\begin{equation}}
\newcommand{\en}{\end{equation}}
\newcommand{\ie}{{\it i.e.}}
\def\CC{I\!\!\!\!C}
\def\cA{{\cal A}}
\def\cD{{\cal D}}
\def\cM{{\cal M}}
\def\cO{{\cal O}}
\def\cW{{\cal W}}
\def\cY{{\cal Y}}

\def\sst{\scriptscriptstyle}

\def\ds{\displaystyle}
\def\Tr{{\rm Tr}}
\def\sun{{\rm SU}(N)}
\def\spn{{\rm Sp}(N)}
\def\son{{\rm SO}(N)}
\def\topic#1{\vspace{1.3cm}
             \onward
             \noindent {{\large \bf \thesection. #1}}
             \vspace{0.3cm}}
\def\Eu{\hchi}

\def\sR{{\sst R}}

\begin{document}
\setlength{\unitlength}{0.25cm}
\thispagestyle{empty}
\hfill              \begin{tabular}{l} {\bf hep-th/9305097} \\
                                        {\sf BRX-TH--346} \\
                                        {\sf JHU-TIPAC--930015} \\
                                     \end{tabular}

\vspace{1.7cm}

\begin{center}
\begin{tabular}{c}
{\LARGE Two-dimensional Yang-Mills Theories } \\[0.2cm]
{\LARGE Are String Theories }
\end{tabular}
\vspace{1.5cm}

\setcounter{footnote}{1}
{\large Stephen G. Naculich,\footnote{Supported in part by
the NSF under grant PHY-90-96198
and by the Texas National Research Laboratory Commission
under grant RGFY-93-292. Address after Sept.~1993:
Department of Physics and Astronomy, Bowdoin College, Brunswick, ME  04011}
 Harold A. Riggs,\footnote{Supported in part by  the DOE under grant
               DE-FG02-92ER40706}
 and Howard J. Schnitzer$^\ddagger$}
\vspace{0.5cm}
{ \normalsize \sl
\begin{tabular}{ll}
\kern 0.0em \begin{tabular}{c}
$^{\dagger}$ Dept. of Physics and Astronomy \\
 The Johns Hopkins University  \\
 Baltimore, MD 21218   \\[0.1cm]
naculich@fermi.pha.jhu.edu
\end{tabular}    &     \begin{tabular}{c}
                          $^\ddagger$Department of Physics \\
                                 Brandeis University  \\
                               Waltham, MA 02254    \\[0.1cm]
                             hriggs, schnitzer@binah.cc.brandeis.edu
                            \end{tabular}
\end{tabular} }\\[1.0cm]

{\normalsize \sf May 1993}

\end{center}

\vfill
\begin{center}
{\sc Abstract}
\end{center}

\begin{quotation}

We show that two-dimensional $\son$ and $\spn$ Yang-Mills
theories without fermions
can be interpreted as closed string theories.
The terms in the $1/N$ expansion of the partition function
on an orientable or nonorientable manifold $\cM$
can be associated with maps from a string worldsheet
onto $\cM$.
These maps are unbranched and branched covers of $\cM$
with an arbitrary number of infinitesimal worldsheet
cross-caps mapped to points in $\cM$.
These string theories differ from $\sun$ Yang-Mills
string theory in that they involve odd powers
of $1/N$ and require both orientable and
nonorientable worldsheets.

\end{quotation}
\vfill

\setcounter{page}{0}
\newpage
\setcounter{page}{1}
\setcounter{section}{0}

\topic{Introduction}

Calculating hadron physics directly from QCD
remains tantalizingly elusive after nearly two decades of effort.
The strong interactions exhibit stringy
characteristics at low momentum transfers
and this provided the original impetus for the
development of string theory as a theory of hadrons.
Formulating QCD as a string theory\cite{thooft,barIbar}
would be an important step in connecting it to hadron physics
and, perhaps, in explaining confinement.

In a series of ground-breaking papers,\cite{gross}$^{-}$\cite{omega}
Gross, Taylor, and Minahan have made progress in this direction by
arguing that two-dimensional QCD is a string theory.
This identification was possible
because the partition function of Yang-Mills gauge theory
on an arbitrary two-dimensional manifold
is known in closed form.\cite{rusakov,migdal}
On a compact orientable surface $\cM_G$ with area $A$ and genus $G$
(so that $\Eu =2-2G$ is the Euler characteristic),
the partition function is given by
\begin{eqnarray}
Z_{\cM_G}
& = &
\int [\cD A^\mu]  \exp
\left[-\,\frac{1}{4 \gauge^2} \int_{\cM_G} {\rm d}^2x \sqrt{g}
 \, \Tr \, F_{\mu\nu} F^{\mu\nu}  \right] \nonumber \\ [.1in]
& = &
\sum_R (\dim R)^{2-2G}  \: \ex^{-\lambda A C_2(R)/2N} \hquad .
\label{orientpart}
\end{eqnarray}
The sum runs over all irreducible representations
$R$ of the gauge group, $\dim R$
and $C_2(R)$ denote the dimension and quadratic Casimir of $R$,
and $\lambda = \gauge^2 N$,
where $\gauge$ is the gauge coupling constant.
The claim that two-dimensional $\sun$ gauge theory without fermions
is equivalent to a closed string theory
(with string coupling $1/N$ and string tension $\lambda/2$)
is verified by relating the expansion of (\ref{orientpart})
in powers of $1/N$ to the genus expansion of a string
theory.\cite{gross}$^{-}$\cite{cdmp}
The terms in the expansion of the partition function
can be interpreted as a weighted counting of
maps from  (possibly disconnected) orientable worldsheets $\cW$
onto the target space $\cM_G$.
The power of $N$ gives the Euler characteristic of the worldsheet,
and the coefficients are related to the number of
inequivalent maps from $\cW$ to $\cM_G$.
The string theory is therefore determined by the
set of worldsheets to be included in its genus expansion,
the types of maps counted,
and the weights of those maps
in the expansion of the $\sun$ Yang-Mills partition function.

In this paper we show that two-dimensional Yang-Mills theories
with gauge groups $\son$ and $\spn$
can also be interpreted as closed string theories.
A significant difference between these theories
and $\sun$ Yang-Mills theory emerges immediately.
The $1/N$ expansion of the partition function for $\sun$
involves only even powers of $1/N$,
whereas we shall see that
the partition functions for $\son$ and $\spn$
necessarily involve odd powers of $1/N$.
In a string interpretation,
these terms correspond to worldsheets of odd Euler characteristic.
Since we are considering gauge theories without fermions,
these worldsheets do not have boundaries
and must necessarily be nonorientable.
Thus the closed string theories
corresponding to $\son$ and $\spn$ Yang-Mills theory
include both orientable and nonorientable worldsheets.

At first sight, the necessity for nonorientable worldsheets
would appear to pose difficulties
for a string interpretation of
$\son$ and $\spn$ Yang-Mills theories
on an orientable surface,
since there are no coverings of orientable target spaces
by nonorientable surfaces\cite{zvc}.
As we will see, however, the odd $1/N$ terms
in the partition function for orientable surfaces
are associated not with true coverings, but with {\em pinch maps},
for which the theorem just cited does not hold.
Thus, we are able to give a consistent string interpretation
of (\ref{orientpart}) for $\son$ and $\spn$ on $\cM_G$.

The $\son$ and $\spn$ Yang-Mills theories on a
{\em nonorientable} surface can also be given
a string interpretation.
An arbitrary compact nonorientable two-manifold
can be constructed as the connected sum of $q$ projective planes,
or as a sphere with the insertion of $q$ cross-caps.
(A cross-cap is a projective plane with a disc removed.)
The resulting surface $\cM_q$,
where $q$ is referred to as the genus,
has Euler characteristic
\eq
\Eu = 2-q \hquad .
\label{euler}
\en
The partition function
\setcounter{footnote}{1}
of Yang-Mills theory on $\cM_q$ is given by\footnote{
The subtleties of quantizing and solving a gauge theory
on a nonorientable manifold
are explained in ref.~\dcite{witten}.}
\eq
Z_{\cM_q} = \sum_{R=\overline{R}} (\eps_{\sR} \dim R)^{2-q}  \:
\ex^{-\lambda A C_2 (R)/2N} \hquad ,
\label{norientpart}
\en
where the sum only runs over self-conjugate representations
($R= \overline{R}$) of the gauge group, and
where $\eps_{\sR} = 1(-1)$
if there exists a symmetric (anti-symmetric) invariant in
$R \otimes R \rightarrow \CC $.

We will demonstrate that
all the leading terms in the $1/N$ expansion of
the $\son$ and $\spn$ Yang-Mills partition functions
on an arbitrary surface $\cM$
(either orientable or nonorientable)
can be associated with surface maps from
orientable and nonorientable worldsheets $\cW$
onto the target space $\cM$.
In general, these maps are compositions of
branched coverings of the target space
with pinch maps.
The pinch maps send infinitesimal cross-caps
on the worldsheet to points on
the target space.
When the Euler characteristic of the target space
satisfies $\Eu \neq 0$,
the $1/N$ expansion contains subleading terms
whose geometric interpretation remains obscure.
For the torus and Klein bottle ($\Eu =0$), however,
the subleading terms vanish,
and all the terms in the partition function
can be given a geometric interpretation.

In section 2,
we present the $1/N$ expansion of the Yang-Mills partition
function for $\son$ and $\spn$.
In section 3,
we clarify the nature of the
Young tableau transposition symmetry
of the $\sun$ Yang-Mills theory\cite{gross}
that guarantees that only even powers of $1/N$
appear in the partition function.
The lack of a similar symmetry for $\son$ and $\spn$
leads to odd powers of $1/N$ in the partition functions
of these theories.
However, certain odd terms are absent in the partition functions
of all three theories due to the presence
of a partial transposition symmetry,
and this fact has a common string interpretation
as the evenness of the number of branch points.
In sections 4 and 5, we
present the string interpretation of $\son$ and $\spn$
Yang-Mills theories.

Intuitively it is easy to see why odd powers of $N$ occur
for $\son$ and $\spn$.
In the double line picture,\cite{thooft}
a closed gluon propagator contains both a ribbon and a
M\"obius band, due to the self-conjugacy of the
fundamental representations of $\son$ and $\spn$,
and thus gives contributions proportional to both $N^2$ and $N$.
In contrast, the closed gluon propagator in $\sun$
contains only a ribbon in the double line formalism,
because the fundamental representation of $\sun$ is not
self-conjugate, so that only even powers of $N$ appear.

\topic{The $1/N$ Expansion of the Yang-Mills Partition Function}

In this section we formulate the $1/N$ expansion of the
partition function for $\son$ and $\spn$ Yang-Mills theory
on an orientable or nonorientable surface.
We begin by expressing the quadratic Casimirs and dimensions
of the representations of $\son$ and $\spn$ as polynomials in $N$.

All irreducible representations of $\spn$
and all irreducible tensor representations of $\son$
can be represented by Young tableaux with at most
$n$ rows, where $n$ is the rank of the group.
(For $\spn$, $N$ is even, and the rank is $n={1\over 2} N$.
For $\son$, the rank $n$ is the integer part of ${1\over 2} N$.)
We will denote the $i^{{\rm th}}$ row length of the tableau by $\ell_i$
so that $\ell_i \geq \ell_{i+1}\geq 0$,
and the $j^{{\rm th}}$ column length by $\col_j$
so that $\col_j \geq \col_{j+1} \geq 0$.
(For the relation between row lengths and
Dynkin indices, see, for example, ref.~\dcite{dual}.)
We denote by
\eq
r = \sum_{i=1}^n \: \ell_i = \sum_{j=1}^{\ell_1} \: \col_j
\en
the number of cells (or boxes) in the tableau
associated with a given representation.

The quadratic Casimirs for $\son$ and $\spn$ are given by
\eq
C_2 (R) = fN \left[ r  + {T(R) \over N} - {\sign r \over N} \right]
\label{casimir}
\en
\setcounter{footnote}{1}
where the long roots of each group satisfy $(\alpha ,\alpha ) =2$,
and where\footnote{$T(R)$ is the quantity denoted
$\tilde{C}(R)$ in refs.~\dcite{gross} and~\dcite{gtaylor},
$\tilde n$ in ref.~\dcite{minahan}, and $X+r$ in ref.~\dcite{dual}.}
\eq
T(R) = \sum^n_{i=1} \, \ell_i (\ell_i + 1 - 2i)
     = \sum^{k_1}_{i=1} \, \ell^2_i - \sum_{j=1}^{\ell_1} k^2_j   \qquad
\label{transquant}
\en
with
$$
\begin{array}{ccl}
f  =1, \qquad &    \sign = ~1, \qquad
&{\rm for} \quad \son, \\
f = {\textstyle {1\over 2}}, \qquad & \sign = -1, \qquad
&{\rm for} \quad\spn.
\end{array}
$$
The permutation sign $\eps_\sR$ is given by
\eq
\eps_\sR =  \sign^r
\label{permsign}
\en
for tensor representations of $\son$ and
all representations of $\spn$.\footnote{$\eps_\sR$
equals the sign in eq.~2.20 of ref.~\dcite{tetra},
where an extensive discussion of these signs may be found.}

The dimension of a representation of $\sun$ corresponding
to a Young tableau $R$ is conveniently given by Robinson's
celebrated hook length formula,\cite{robinson}
\eq
       (\dim R)_{\sun} = \prod_{x\in R} {N + a(x) \over h(x)} \hquad .
\label{robhook}
\en
The product in (\ref{robhook}) runs over all cells $x$ of $R$,
each of which is identified by its row $i$ and column $j$.
The {\em hook length} of a cell is given by
\eq
h(x) = h(i,j) = \ell_i + k_j -i-j+1
\en
and $a(x) = a(i,j) = j-i$.
The hook length product is related to the dimension
$d_\sR$ of the representation of the {\em symmetric} group $S_r$
specified by the tableau $R$ by\cite{frt}
\eq
  \ds    {1\over \ds \prod_{x \in \sR} h(x)} =  {d_\sR \over r!}  \hquad.
\en
The analogs of Robinson's formula for $\son$ and $\spn$ are\cite{tabking}
\eq
(\dim R)_{\spn} = \prod_{x\in \sR} \: \frac{N - c(x)}{h(x)}
\en
and
\eq
(\dim R)_{\son} = \prod_{x\in \sR} \: \frac{N + s(x)}{h(x)}
\en
where
\eq
c(x) = c(i,j) = \left\{
\begin{array}{ll}
k_i + k_j - i - j, & \qquad i \leq j, \\
-\ell_i - \ell_j + i + j -2, & \qquad i > j,
\end{array}
\right.
\en
\eq
s(x) = s(i,j) = \left\{
\begin{array}{ll}
\ell_i + \ell_j - i - j, & \qquad i \geq j, \\
-k_i - k_j + i + j -2, & \qquad i < j.
\end{array}
\right.
\en
A straightforward calculation gives
\eq
\sum_{x\in \sR} a(x) = - \sum_{x\in \sR} c(x) = \sum_{x\in \sR} s(x)
  = {\textstyle \frac{1}{2}} T(R).
\en
The hook length formulae are very convenient for computing
the $1/N$ expansion of the dimension of $R$.
The leading terms of this expansion for representations of
$\sun$, $\son$, and $\spn$ are given by
\eq
\dim R  = \frac{d_\sR N^r}{r!} \:
\left[ 1 + \frac{T(R)}{2N} + \cO ({1 \over N^2}) \right]  ,
\label{dimexpand}
\en
where the $\cO(1/N^2)$ terms are different for each group.

We now have the ingredients to write
the $1/N$ expansion of the partition function
(\ref{orientpart}) or (\ref{norientpart}) on a surface $\cM$. {}From
(\ref{casimir}), we see that tableaux with $r \ll N$
have quadratic Casimirs of $\cO (N) $,
and so will contribute perturbatively to the partition function.
It is straightforward to prove that if $r$ is {\em not} $ \ll N$,
then the Casimir of the representation is of $\cO(N^2)$ or higher.
Spinor representations of $\son$ also have Casimirs $ \geq \cO(N^2)$.
Consequently, the contribution of these representations
to the partition function is exponentially suppressed,
and corresponds to effects non-perturbative in $1/N$.
(This is in contrast to $\sun$, where certain representations
with $\cO(N)$ boxes,
namely, {\em composite} representations\cite{gtaylor},
have Casimirs of $\cO(N)$,
and so contribute perturbatively.)
We will neglect non-perturbative effects in this paper,
and consider only Young tableaux with $ r \ll N $,
denoting the set of such tableaux by $\cY$.
Let $\cA = f \lambda A$ denote the dimensionless area of the surface $\cM$.
Using (\ref{casimir}), (\ref{permsign}), and (\ref{dimexpand}),
we write the perturbative Yang-Mills partition function
on an orientable (\ref{orientpart}) or a nonorientable (\ref{norientpart})
two-manifold $\cM$ with Euler character $\Eu$ as
\begin{eqnarray}
Z_{\cM}
& = & \sum_{\sR\in\cY}
(\sign^r \dim R)^{\Eu}
\; \ex^{-\cA r/2} \; \ex^{\sign \cA r/2N}  \; \ex^{-\cA T(R)/2N}  \\
& = & \sum_{\sR\in\cY} \sum_{i=0}^\infty
\left(\sign^r N^r d_R \over r! \right)^{\Eu}  \;
{1 \over i!} \left(-\cA T(R)\over 2N\right)^i
 \ex^{-\cA r/2} \; \ex^{\sign \cA r/2N}
\left[ 1 + \cO({1 \over N}) \right]  .\qquad
\end{eqnarray}
(All the tableaux in $\cY$ are self-conjugate,
and so contribute to the partition function on nonorientable surfaces.)
The subleading terms in the square brackets
result solely from the subleading contributions
to the dimension of $R$
in (\ref{dimexpand}), and have no dependence on $\cA$.

For each of the tableaux $R \in \cY$,
there exists a transposed tableau $\tR$, also in $\cY$,
obtained by interchanging rows and columns.  {}From
(\ref{transquant}), we see that\cite{ns,dual,gross}
\eq
T(\tR) = - T( R ).
\label{transsym}
\en
Rewriting $ \sum_R = {1\over 2} \left[ \sum_R + \sum_{\tR} \right]$
and using (\ref{transsym}), we find
\eq
 Z_{\cM} =
\sum_{\sR\in\cY} \sum_{i=0}^{\infty}
\left( \sign^r N^r d_R \over r! \right)^{\Eu}  \;
{1 \over (2i)!} \left(\cA T(R)\over 2N\right)^{2i}
\ex^{-\cA r/2} \; \ex^{\sign \cA r/2N}
\left[ 1 + \cO({1\over N^2}) \right]  .
\label{genform}
\en
Now the subleading terms in the square brackets are all of the form
$\cA^m/N^n$ with $m < n$.
They are absent
in the case of Yang-Mills theory on a torus or Klein bottle ($\Eu=0$).

In sections 4 and 5, we will interpret the leading order terms
in (\ref{genform})
 ({\em all} the terms in the case of the torus or Klein bottle)
as weighted multiplicities of maps from (possibly disconnected)
orientable and nonorientable worldsheets onto the target space $\cM$.

\newpage
\topic{Young Tableau Transposition Symmetry}

Before presenting the string interpretation of the partition function,
we briefly comment on a transposition symmetry present in Yang-Mills theory.
In $\sun$ Yang-Mills theory, there is a symmetry under which
\eq
C_2 (\pT, N) = - C_2 (T, -N), \qquad
\Bigl| \dim (\pT, N) \Bigr| = \Bigl| \dim (T, -N) \Bigr|, \qquad
{\rm for} \quad \sun
\label{symmetry}
\en
where $T$ is the {\em composite } representation  ${\bar S} R$
(see ref.~\dcite{gtaylor}), and $\pT = {\overline \tS} \tR $.
This symmetry was noted in ref.~\dcite{gross}
in the case of {\em chiral} representations ($S=1$),
but it is easy to prove that the more general symmetry
(\ref{symmetry}) holds for representations of $\sun$.
Since both $T$ and $\pT$ contribute perturbatively to
the partition function sum,
this symmetry guarantees the absence of odd powers of $1/N$
in the partition function.  As noted in the introduction,
this means that $\sun$ requires only orientable worldsheets.
The symmetry (\ref{symmetry})
also ensures that only even powers of $ \cA T(R)/N$ appear,
which in the string interpretation is related
to the evenness of the number of branch
points on a surface without boundary.

In the case of $\son$ and $\spn$, the analogous symmetry for the Casimirs
does {\em not} hold because of the last
term in (\ref{casimir}).
While (\ref{dimexpand}) shows that
\eq
 \dim (\tR, N)   =  (-1)^r \dim (R, -N)
\qquad {\rm to} \quad \cO(1/N),
\en
for $\son$ and $\spn$ representations,
this transposition symmetry breaks down at $\cO(1/N^2)$,
as seen in the example
\eq
    (\dim \, \young4000001\, )_{\spn} =
  {N^4 \over 4!} \left( 1 + {6\over N} + {11\over N^2} + {6 \over N^3}\right),
\en
\eq
  (\dim \, {\raise 0.375cm \hbox{\young1111004}}\, )_{\spn} =
  {N^4 \over 4!} \left( 1 - {6\over N} - {1\over N^2} + {6 \over N^3}\right).
\en
As a result of the breakdown of this symmetry,
the partition function contains odd powers of $1/N$,
and therefore implies the existence of nonorientable worldsheets.

However, the transposition symmetry (\ref{transsym})
that does remain valid in the case of $\son$ and $\spn$
has the effect that only {\em even} powers of $\cA T(R) /N$
appear in the leading order terms of (\ref{genform}).
In the following section, we interpret
these terms as arising from branched coverings with simple
(order two) branch points,
and it is a basic
topological fact that the number of such branch points is even
on any surface without boundary.  {}From
the vantage point of Yang-Mills theory,
the transposition symmetry (\ref{transsym}) enforces
this evenness, and thus makes possible for all three groups
the string theory interpretation in terms of branched coverings.

\newpage
\topic{Nonorientable Target Spaces}

In this section
we show that $\son$ or $\spn$ Yang-Mills theory on a
nonorientable surface $\cM_q$
is equivalent to a closed string theory with target space $\cM_q$.
Each term in the Yang-Mills free energy corresponds
to some surface map from a connected worldsheet $\cW$
onto the nonorientable target space $\cM_q$.
Since the free energy corresponds to the sum over maps
from {\em connected} worldsheets,
the partition function corresponds to the sum over
maps from {\em all} worldsheets, both connected and disconnected.
Inequivalent maps from $\cW$ to $\cM_q$
give distinct contributions to the free energy,
but the precise nature of the equivalence is somewhat unclear.
Topologically inequivalent maps clearly count as distinct.
The presence of geometric moduli, however,
seems to indicate that some topologically  equivalent maps
are also distinguished, and that the equivalence relation
is a refinement of the purely topological classification.
While we intend to return to this question in the near future,
in this paper we will adopt the
combinatorial procedure of ref.~\dcite{gtaylor} to count distinct maps.

We begin by considering (possibly disconnected) $r$-fold
unbranched covers of $ \cM_q $.
They are characterized by the fact that exactly $r$ points of
the cover are mapped to each point of the target space.
The generators $a_j$, $j=1, \ldots, q$,
of the fundamental group $\pi_1 (\cM_q)$ satisfy the single relation
\eq
a_1  a_1 \cdots  a_q a_q = 1 \hquad .          \label{relation}
\en
Let $\nu$ be a (possibly disconnected) $r$-fold unbranched
covering of $\cM_q$.
Choose a point $p$ of $\cM_q$, and label the $r$ sheets
of $\nu$ over $p$ by the integers in $I = \{ 1,2, \ldots , r \}$.
With each element $t \in \pi_1 (\cM_q )$,
we associate the permutation of $I$ that results from the transport
of the labels on sheets around the path obtained by lifting $t$ to $\nu$.
This procedure defines a homomorphism
$H_\nu$ from $\pi_1  (\cM_q )$ to the permutation group $S_r$.
Homeomorphisms of the covering surface can permute the labeling
of the sheets,
but they leave the homomorphism $H_\nu$ invariant.
For a given covering $\nu$,
there are $r!$ different labelings of the sheets.
Relabeling the $r$ sheets with the permutation $\rho$
gives the conjugate homomorphism  $\rho H_\nu \rho^{-1}$.
If $\rho$ belongs to $S_\nu \subset S_r$,
the group of permutations produced by
homeomorphisms of the covering surface,
then $\rho H_\nu \rho^{-1} = H_\nu $.
Thus, the number of {\em distinct}
homomorphisms corresponding to $\nu$ is $r!/|S_\nu |$,
where $|S_\nu|$ is the number of elements of $S_\nu$.

Next consider an $r$-fold {\em branched} covering of
$\cM_q$ with $\nbr$ branch points $b_1, \ldots, b_\nbr$.
We will only need the generic case of {\em simple} branched coverings,
in which exactly $r-1$ points of the worldsheet
are mapped to each branch point on $\cM_q$
and $r$ points are sent to every other point of $\cM_q$.
It is a topological fact that the number of such branch points
is necessarily even.
Choose a set of $\nbr$ curves $\{c_j\}$, each of which
encircles one of the branch points $b_j$.
The $c_j$ together with the $a_j$ form a set of generators for the
fundamental group $\pi_1 (\cM \backslash  \{ b_j \} )$,
defined by the single relation
\eq
c_1 \cdots c_\nbr a_1  a_1 \cdots  a_q a_q  = 1 \hquad .
\label{fundrel}
\en
As before, an $r$-fold covering $\nu$ with branch points
$b_1, \ldots, b_\nbr$
defines a homomorphism
from $\pi_1 (\cM \backslash \{ b_j \} )$ to $S_r$,
but each of the generators $c_j$ is associated with a permutation
$p_j$ belonging to $ P_r$,
the conjugacy class of permutations that interchange only two elements,
since the branched cover is simple.

Let $\Sig (q,r,\nbr)$ denote the set of (connected and disconnected)
$r$-fold covers of $\cM_q$ with $\nbr$ branch points.
We wish to count each covering with a weight of $1/|S_\nu|$.
In light of the previous discussion,
this is equivalent to counting distinct homomorphisms $H_\nu$
with a weight of $(1/r!)$.
The weighted sum over coverings is given by
\eq
\sum_{ \nu\in\Sig(q,r,\nbr) } \; \frac{1}{|S_\nu |}
= \sum_{p_1 \cdots p_\nbr \in P_r} \;
  \sum_{t_1 \cdots t_q \in S_r} \;
\frac{1}{r!} \,
\delta (p_1 \cdots p_\nbr \, t^2_1 \cdots  t^2_q )
\label{weightsum}
\en
where the delta function $\delta  (\rho)$, defined by
\eq
\delta  (\rho ) =
\left\{
\begin{array}{lll}
1 & {\rm if} & \rho ~ = ~{\rm identity}  \\
0 & {\rm if} & \rho ~ \neq ~{\rm identity} ,
\end{array}
\right.
\en
enforces the relation (\ref{fundrel}).
Let $D_R(\rho )$ be the matrix associated with $\rho \in S_r$ in the
representation of $S_r$ specified by the Young tableau $R$.
The character of $\rho$ is given by $\hchi_R (\rho ) = \Tr \, D_R (\rho )$.
By the orthogonality of characters, we have
\eq
\delta  (\rho ) = \frac{1}{r!} \: \sum_R \: d_R \: \hchi_R (\rho )
\en
where $d_R = \Tr ~I_R $ is the dimension of $R$.
($I_R$ is the identity matrix in the representation $R$.)
Thus (\ref{weightsum}) can be expanded as
\eq
\sum_\nu \: \frac{1}{|S_\nu |}
= \sum_{p_1 \cdots p_\nbr \in P_r} \; \sum_{t_1\cdots t_q \in S_r} \;
\sum_R \, \left( \frac{1}{r!} \right)^2
d_R \, \hchi_R (p_1 \cdots p_\nbr t^2_1 \cdots t^2_q ) \; .
\label{weightexpand}
\en
To compute this, consider
\begin{eqnarray}
\lefteqn{ \sum_{p_1 \cdots p_\nbr \in P_r} \;
\sum_{t_1 \cdots t_q \in S_r} \;
D_R(p_1  \cdots p_\nbr t^2_1 \cdots t^2_q) }\nonumber \\ [.1in]
 &=&   \sum_{p_1 \in P_r}    \, D_R (p_1)
\cdots \sum_{p_\nbr \in P_r} \, D_R (p_\nbr)
       \sum_{t_1 \in S_r} \, D_R (t^2_1 )
\cdots \sum_{t_q \in S_r} \, D_R (t^2_q )  \hquad .
\end{eqnarray}
Using Schur's lemma and the fact that all the representations
of $S_r$ are real, it follows that
\eq
\sum_{t \in S_r} \, D_R(t^2) = \frac{r!}{d_R} \: I_R.
\en
In ref. \dcite{gtaylor}, it is shown that
\eq
\sum_{p\in P_r} \; D_R(p)
= {T(R) \over 2}  I_R.
\en
Therefore,
\eq
\sum_{p_1 \cdots p_\nbr \in P_r} \;
\sum_{t_1 \cdots t_q \in S_r} \;
D_R (p_1 \cdots p_\nbr \, t^2_1 \cdots t^2_q)
= \left( \frac{T(R)}{2} \right)^\nbr
\; \left( \frac{r!}{d_R} \right)^q \; I_R.
\en
Thus, the weighted sum (\ref{weightexpand})
over the coverings of $\cM_q$ is finally given by
\eq
\sum_{\nu\in\Sig(q,r,\nbr)} \; \frac{1}{|S_\nu |}
= \sum_R \:
\left( \frac{T(R)}{2} \right)^\nbr
\left( \frac{r!}{d_R} \right)^{q-2} .
\label{finalsum}
\en
The representations $R$ summed over in (\ref{finalsum}) correspond
to the set of Young tableaux with $r$ boxes.

Using (\ref{finalsum}), we may rewrite the
$\son$ and $\spn$ Yang-Mills partition function on $\cM_q$
(\ref{genform})
as a sum over branched coverings of $\cM_q$
\eq
Z_{\cM_q}
= \sum_{r=0}^{\infty}
  \sum_{i=0}^{\infty}
  \sum_{\nu\in\Sig(q,r,2i)} \;
\frac{1}{|S_\nu|} \; \sign^{(2-q)r}
\; \ex^{-\cA r/2} \; \ex^{ \sign \cA r / 2N }
{\cA^{\nbr} \over (\nbr)!} \left( 1 \over N \right)^{(q-2)r + \nbr}
\left[ 1 + \cO({1 \over N^2}) \right]  .
\label{coversum}
\en
All the leading terms in the partition function (\ref{coversum})
can be interpreted in terms of
surface maps from worldsheets onto $\cM_q$, as follows.

The leading $1/N$ term
\eq
   \sum_{r=0}^{\infty} \sign^{(2-q)r} \ex^{-\cA r/2}
  \sum_{\nu\in\Sig(q,r,0)} \;
\frac{1}{|S_\nu|} \;  N^{r (2-q) }  \hquad ,
\label{areaindterm}
\en
corresponds to a sum over
unbranched $r$-fold coverings (local homeomorphisms).
For such coverings, we necessarily have\cite{zvc}
$$
 \Eu_{\cW} = r \Eu_{\cM_q} = r (2-q)  \hquad ,
$$
and the area of the worldsheet is $r \cA$,
so these maps contribute with a factor $N^{(2-q)r}$
and with the action given by the string coupling
times the area of the worldsheet.
The coefficient of this term is the weighted sum over coverings
$ \sum_{\nu \in \Sig (q,r,0)}  1/|S_\nu | $.
The covering spaces can be either connected or disconnected,
and either orientable or nonorientable
(but must be nonorientable when $(2-q)r$ is odd).
For example, consider coverings of the Klein bottle ($q=2$).
The weighted sum over $r$-fold coverings is given by
\eq
 \sum_{\nu \in \Sig (q=2,r,0)}  {1\over |S_\nu|} = \sum_R (1) = p(r) \hquad ,
\label{kleincover}
\en
where $p(r)$ is the number of partitions of $r$.
Using the relation between the partition function and the
free energy discussed above,
the number of {\em connected} $r$-fold coverings of the Klein bottle is
therefore $q(r)= \sum_{ d|r} d$, the sum of divisors of $r$.
If $r$ is odd, the covering surface must be nonorientable,
\ie, a Klein bottle.
If $r$ is even, however, the Klein bottle can also be covered
by the torus.
In fact, for even $r$,
we count $q(r/2)$ connected $r$-fold
coverings of the Klein bottle by the torus,
and $q(r)-q(r/2)$ connected $r$-fold coverings by the Klein bottle,
using a procedure analogous
to that described in ref.~\dcite{minahan}.

Terms in (\ref{coversum}) with $i \neq 0$
correspond to branched coverings.
The addition of each pair of branch points decreases
the Euler characteristic of the worldsheet by two,
so that the Euler
characteristic of the world sheet is $(2-q)r - \nbr$ for
an $r$-fold branched covering with $\nbr$ branch points.
Integrating over the positions of $\nbr$ branch points
on the target space and accounting for their indistinguishability
gives the factor $\cA^{\nbr}/(\nbr)!$.

Finally, the sum (\ref{coversum}) contains a factor
\eq
 \exp \left( \sign \cA r \over 2N \right)
= \sum_{k=0}^\infty  {1\over k!} \left( \sign \cA r \over 2N \right)^k \hquad.
\label{crosscaps}
\en
The $k^{{\rm th}}$ term in this sum corresponds to
an $r$-fold covering of $\cM_q$ with $\nbr$ branch points
and with the insertion of $k$ infinitesimal cross-caps.
The insertion of one or more cross-caps renders the
worldsheet nonorientable.
The surface maps from these worldsheets are compositions
of unbranched or branched coverings with  {\em pinch maps},
in which the cross-caps on the worldsheet are mapped to
points on the target space.
Each cross-cap decreases the Euler characteristic of the worldsheet
by one,
contributing an overall factor $1/N^k$.
Integrating the positions of the $k$ cross-caps over the area
of the worldsheet,
and taking account of their indistinguishability,
gives the factor $(r \cA)^k/k!$.  {}From (\ref{crosscaps}) we see that
each cross-cap contributes a factor ${1 \over 2}$ for $\son$
and  $-{1 \over 2}$ for $\spn$.

We have exhibited all the leading terms as
a genus expansion of weighted multiplicities of surface maps.
The remaining $\cO(1/N^2)$ terms in (\ref{coversum}),
for which we cannot yet account,
are due to the subleading corrections to the dimension of $R$.
For Yang-Mills theory on a Klein bottle these subleading terms vanish,
and we have succeeded in giving a string interpretation for
all the terms in the partition function.

\vspace{0.3cm}
\topic{Orientable Target Spaces}

In this section, we show that there appears to be
no obstacle to a string theory interpretation
of $\son$ or $\spn$ Yang-Mills theory on an {\em orientable} surface,
analogous to that for nonorientable surfaces given in the last
section.

It was shown in ref.~\dcite{gtaylor}
that the weighted sum over $r$-fold coverings of
an  orientable genus $G$ target space $\cM_G$
with $\nbr$ branch points is given by
\eq
\sum_{\nu\in\Sig(G,n,\nbr)} \; \frac{1}{|S_\nu |}
= \sum_R \:
\left( \frac{T(R)}{2} \right)^\nbr
\left( \frac{r!}{d_R} \right)^{2G-2} .
\en
Using this, the partition function on $\cM_G$ (\ref{genform})
may be written as a weighted sum over coverings of $\cM_G$
\eq
Z_{\cM_G}
= \sum_{r=0}^{\infty}
  \sum_{i=0}^{\infty}
  \sum_{\nu\in\Sig(G,r,2i)} \;
\frac{1}{|S_\nu|} \;
\ex^{-\cA r/2} \;
\ex^{ \sign \cA r / 2N }
{\cA^{\nbr}  \over (\nbr)!} \left(1 \over N \right)^{(2G-2)r+\nbr}
\left[ 1 + \cO({1 \over N^2}) \right]  .
\en
As before, the leading terms in this partition function
can be interpreted in terms of
surface maps from worldsheets onto $\cM_G$.

The leading $1/N$ term corresponds
to a weighted sum over unbranched $r$-fold coverings of $\cM_G$.
In contrast to the last section,
the covering spaces here must be orientable\cite{zvc}.
For example, the weighted sum over $r$-fold coverings of the torus is
\eq
 \sum_{\nu \in \Sig (G=1,r,0)}  {1\over |S_\nu|} = \sum_R (1) = p(r) \hquad .
\label{toruscover}
\en
Reasoning as before,
the number of {\em connected} $r$-fold coverings of the torus is $q(r)$,
and the covering spaces here are all tori.

Terms with $i \neq 0$ correspond to branched coverings,
and the counting is the same as before.
We again interpret the term $ \exp(\sign \cA r/ 2N )$
as due to surface maps constructed by
inserting infinitesimal cross-caps into
(possibly branched) covering spaces.
This necessarily renders the worldsheet nonorientable.
Note that such maps from nonorientable worldsheets
onto orientable target spaces do exist,
because the maps are not true covers
but compositions of covering maps with  pinch maps.
Therefore, there appears to be no reason to exclude them
from the genus expansion of an unoriented string,
and indeed they make possible a string interpretation of
$\son$ and $\spn$ gauge theories on orientable target spaces,
including the physical case of the torus.

\vspace{0.3cm}
\topic{Concluding Remarks}

We have shown that two-dimensional $\son$ and $\spn$ Yang-Mills
theories without fermions can be understood as closed string theories.
All the leading terms in the $1/N$ expansion of the partition function
on a manifold $\cM$
can be interpreted in terms of maps from a string worldsheet
onto a target space $\cM$.
These maps include unbranched and branched coverings of $\cM$
with an arbitrary number of
infinitesimal worldsheet cross-caps mapped to points in $\cM$.

These string theories differ from $\sun$ Yang-Mills
string theory in that the worldsheets need not be orientable.
In particular, terms in the expansion of the partition function
with odd powers of $1/N$ necessarily correspond to nonorientable worldsheets.

It is intriguing that although the (perturbative)
Yang-Mills partition function
has exactly the same form on an orientable and a nonorientable
manifold with the same Euler characteristic,
the terms have different string theory interpretations.
For example, the leading $1/N$ term
of the partition function on the Klein bottle (\ref{kleincover})
corresponds to maps from the Klein bottle and torus to the Klein bottle,
whereas the identical term in the partition function on the torus
(\ref{toruscover})
corresponds to maps from the torus to the torus.

\vspace{0.5cm}
\noindent {\bf Acknowledgement}
\vspace{0.2cm}

\noindent We thank Daniel Ruberman for several
helpful discussions.



\begin{thebibliography}{99.}
\bibitem{thooft} G. 't Hooft, \sl Nucl. Phys. \bf B72 \rm (1974) 461;
                                               \bf B75 \rm (1974) 461
\bibitem{barIbar} W. Bardeen, I. Bars, A. Hanson, and R. Peccei,
                 \sl Phys. Rev. \bf D13 \rm (1976) 2364
\bibitem{gross}  D. Gross, `Two-Dimensional QCD As a String Theory,'
                 PUPT-1356, LBL-33415, hepth-9212149, December 1992
\bibitem{minahan}J. Minahan, `Summing Over Inequivalent Maps in the
                 String Theory Interpretation of Two-Dimensional QCD,'
                 UVA-HET-92-10, hepth-9301003, December 1992
\bibitem{gtaylor} D. Gross and W. Taylor, `Two-Dimensional QCD Is a String
                 Theory,' PUPT-1376, LBL-33458, hepth-9301068, January 1993
\bibitem{omega}  D. Gross and W. Taylor, `Twists and Wilson Loops in the
                 String Theory of Two Dimensional QCD,'
                 PUPT-1382, LBL-33767, hepth-9303076, March 1993
\bibitem{minpol} J. Minahan and A. Polychronakos, `Equivalence of Two
                 Dimensional QCD and the $c=1$ Matrix Model,'
                 UVA-HET-93-02, hepth-9303153, March 1993
\bibitem{douglas}   M. Douglas, `Conformal Field Theory Techniques
                  for Large $N$ Group Theory,' RU-93-13, hepth-9303159,
                  March 1993;  M. Douglas and V. Kazakov,
                  `Large $N$ Phase Transition
                  in Continuum ${\rm QCD}_2$,' RU-93-17, LPTENS-93/20,
                  hepth-9305047, May 1993
\bibitem{cdmp}    M. Caselle, A. D'Adda, L. Magnea, and S. Panzeri,
                 `Two Dimensional QCD is a One Dimensional Kazakov-Migdal
                   Model,' DFTT 15/93, hepth-9304015, April 1993
\bibitem{rusakov}  B. Rusakov, \sl Mod. Phys. Lett. \bf A5 \rm (1990) 693
\bibitem{migdal} A. Migdal, \sl Sov. Phys. JETP \bf 42 \rm (1975) 413
\bibitem{zvc}    H. Zieschang, E. Vogt, and H. Coldewey,
                 \it Surfaces and Planar Discontinuous Groups,
                 \rm Lecture Notes in Mathematics \bf 835 \rm
                 (Springer, New York, 1980)
\bibitem{witten} E. Witten, \sl Comm. Math. Phys. \bf 141 \rm (1991) 153
\bibitem{dual}  S. Naculich, H. Riggs, H. Schnitzer, \sl Phys. Lett.
                \bf B246 \rm (1990) 417
\bibitem{tetra}  S. Naculich, H. Riggs, H. Schnitzer, \sl Nucl. Phys.
                 \bf B394 \rm (1993) 445
\bibitem{robinson} G. Robinson, \it Representation Theory of the
                   Symmetric Group \rm (EUP, Edinburgh, 1961)
\bibitem{frt}  J. Frame, G. Robinson, R. Thrall, \sl Canad. J. Math.
                \bf 6 \rm (1954) 316
\bibitem{tabking}  N. El Samra and R. King, \sl J. Phys. \bf A12 \rm
                 (1979) 2317
\bibitem{ns}  S. Naculich and H. Schnitzer, \sl Nucl. Phys. \bf
              B347 \rm (1990) 687
\end{thebibliography}
\end{document}